\documentclass[12pt]{article}
\usepackage{hyperref}   
\usepackage{latexsym}   
\usepackage{amsmath, amssymb, amsthm}
\usepackage{eufrak}

\newcommand{\real}{\mathbb{R}}
\newcommand{\nature}{\mathbb{N}}
\newcommand{\ra}{\rightarrow}

\newcommand{\e}{\epsilon}

\newcommand{\accu}{{\rm Accu}}
\begin{document}

\newtheorem{thm}{Theorem}[section]
\newtheorem{defi}{Definition}[section]
\newtheorem{fact}{Fact}
\newtheorem{lemma}{Lemma}[section]
\newtheorem{prop}{Proposition}[section]
\newtheorem{coro}{Corollary}[section]
\newtheorem{ex}{Example}[section]
\newtheorem{remark}{Remark}[section]
\newcommand{\rij}[2]{{\bf R}_{#1#2}}
\newcommand{\bprime}{{\prime\prime}}
\newcommand{\tprime}{{\prime\prime\prime}}
\bibliographystyle{plain}   

\title{Multi-agent coordination using nearest-neighbor rules:
revisiting the Vicsek model
\thanks{This work was partly supported by the National Foundation of Natural
Science of China (60305005, 60321002, 60496321) and by a Hong Kong
CERG (Grant No. CityU 1234/03E).}}
\author{Sanjiang Li
\thanks{Email: lisanjiang@tsinghua.edu.cn (S. Li), iswang@cityu.edu.hk (H. Wang)}\\
State Key Laboratory of Intelligent Technology and
  Systems\\
Department of Computer Science and Technology\\
Tsinghua University, Beijing 100084, China\\
 \and
Huaiqing Wang\\
 Department of Information Systems\\
City University of Hong Kong, Hong Kong, China}

\maketitle
\begin{abstract}Recently,
Jadbabaie, Lin, and Morse (2003) offered a mathematical analysis
of the discrete time model of groups of mobile autonomous agents
raised by Vicsek \emph{et al.} in 1995. In their paper, Jadbabaie
\emph{et al.} showed that all agents shall move in the same
heading, provided that these agents are \emph{periodically linked
together}. This paper sharpens this result by showing that
coordination will be reached under a very weak condition that
requires all agents are \emph{finally linked together}. This
condition is also strictly weaker than the one Jadbabaie \emph{et
al.} desired.

\end{abstract}

\section{Introduction}
Coordination of groups of mobile autonomous agents
(\emph{particles} \cite{Vicsek95}, or \emph{boids} \cite{Rey87})
has attracted researchers in a surprisingly wide variety of
disciplines ranging from physics \cite{Vicsek95, Toner95,
Toner98}, to the biological sciences \cite{War,Flierl},  to
computer science and engineering \cite{Rey87,JLM03,Sav04,LBF04}.


This paper is mainly concerned with one particular discrete time
model of groups of mobile autonomous agents, viz., the one
proposed by Vicsek \emph{et al.} \cite{Vicsek95} in 1995. In this
model, a group of autonomous agents is moving in the plane with
all agents moving at the same speed but with different headings.
Each agent's motion is updated using a local rule based on the
average of its own heading and the heading of its ``neighbors."
This is known as the \emph{nearest-neighbor rule} in \cite{JLM03}.
Agent $i$'s \emph{neighbors} at time $t$ are those agents that are
either in or on a circle of pre-specified radius $r$ centered at
agent $i$'s current position. Known as the \emph{Vicsek model},
this can be viewed as a special version of a model proposed by
Reynolds \cite{Rey87} for simulating animal aggregation for the
computer animation industry. Although the Vicsek model is very
simple, simulation results in \cite{Vicsek95} show that, using the
local update rule, all agents shall eventually move in the same
direction despite the absence of centralized coordination, and
that neighborhood of each agent will change.

Recently, Jadbabaie, Lin, and Morse \cite{JLM03} offered a
mathematical analysis of this model and provided a theoretical
explanation for the observed behavior. They adopt a more
conservative approach, which ignores how the neighbor-graphs
depend on the agent positions in the plane. Note that under this
assumption, the Vicsek model is  a graphic example of a switched
linear system. Their goal in that paper was to determine for a
certain large class of switching signals and for any initial set
of agent headings that the headings of all agents will converge
into the same steady heading.

Jadbabaie \emph{et al.} \cite{JLM03} established sufficient
conditions given in terms of neighbor-graphs for coordination of
agents. One main result of \cite{JLM03} shows that all agents
shall eventually move in the same heading if these graph are
\emph{periodically jointly connected}, \emph{i.e.}, the union of
any $T$ sequential graphs is connected for some fixed $T$. This is
a nice result, but as Jadbabaie \emph{et al.} put it \cite[p990,
below Theorem 2]{JLM03}, what one would prefer instead is to show
that coordination would be reached eventually for every switching
signal for which there is an infinite sequence of bounded,
non-overlapping (but not necessarily contiguous) intervals across
which the agents are linked together.

This paper will show above condition desired by Jadbabaie \emph{et
al.} is indeed a sufficient condition for asymptotic convergence.
This, however, follows from a more general observation: our main
result in this paper shows that convergence will be attained if
these neighbor-graphs are \emph{finally jointly connected},
\emph{i.e.}, the union of all graphs started from any time is
connected.

The structure of this paper is as follows. In Section II we give a
formal description of the Vicsek model in terms of switching
signals. Section III provides the major results, for both
leaderless coordination and leader-following coordination.
Conclusions and future work are given in the last section.

\section{The Vicsek model and the nearest-neighbor rule}
In this section, we review some basic definitions concerning the
Vicsek model.

The system studied by Vicsek \emph{et al.} \cite{Vicsek95}
consists of $n$ autonomous agents, e.g., particles, robots, etc.,
labeled 1 through $n$. All agents move in the plane with the same
speed but with different headings. The system operates at discrete
time $t=0,1,2,\cdots$ Let $r>0$ and $v>0$ be given numbers
associated with the system. The dynamics of agent $i$ is described
by the sequence $\{x_i(t),y_i(t),\theta_i(t)\}$, where
$x_i(t),y_i(t)\in\real$ are the coordinates of the agent in the
plane, and $\theta_i(t)$ is its heading taking value from
$[0,2\pi)$. At any time $t=0,1,2,\cdots$, each agent's heading is
updated using a simple rule based on the average of its own
heading plus the headings of its neighbors. For any two agents
$i,j$, we say $j$ is a \emph{neighbor} of $i$ at time $t$, written
$j\in\mathcal{N}_i(t)$, if $d(i,j)\leq r$,\footnote{In certain
situation, choosing open neighborhood would give rise to more
desirable results, see \cite{Jad03}.} where
\begin{equation}
d(i,j)=\sqrt{(x_i(t)-x_j(t))^2+(y_i(t)-y_j(t))^2}.
\end{equation}
Then, agent $i$'s next heading is defined as
\begin{equation}\label{eq:lur}
\theta_i(t+1)=\frac{\theta_i(t)+\Sigma_{j\in\mathcal{N}_i(t)}\theta_j(t)}{1+n_i(t)}
\end{equation}
where $n_i(t)$ is the number of agent $i$'s neighbors at time $t$.
Agent $i$'s next coordinates are defined as
\begin{eqnarray}\label{eq:coor}
  x_i(t+1) &=& x(t)+v_i(t)\cos(\theta_i(t)) \label{eq: x-coor}\\
  y_i(t+1) &=& y(t)+v_i(t)\sin(\theta_i(t)). \label{eq: y-coor}
\end{eqnarray}

For any time $t\geq 0$, we define the \emph{neighbor-graph} of the
system described above as the simple undirected graph
$\mathbb{G}(t)$ over $V=\{1,2,\cdots,n\}$ where the vertex $i$
corresponds to agent $i$ and two vertexes, $i,j$, are connected by
an edge in the graph if  they are neighbors at time $t$,
\emph{i.e.}, if $j\in\mathcal{N}_i(t)$. Since the neighbor
relation can change over time, so can the graph that describes
them. In the sequel, we write $\mathcal{P}$ for the collection of
simple undirected graphs over $V$. A \emph{switching signal} is a
function $\sigma:\nature\ra\mathcal{P}$ that assigns to each time
$t$ a neighbor-graph that specifies the neighbor relation between
agents. Clearly, for a Vicsek model, the function that assigns to
each time $t$ the neighbor-graph $\mathbb{G}(t)$ is a switching
signal.

Note that for the Vicsek model, the neighbor-graph is determined
by the initial positions and headings of all agents as well as the
pre-specified $r>0$ and $v>0$. A complete description of the model
would have to explain explicitly how $\sigma$ changes over time.
As it is difficult to take this into account in a convergence
analysis, Jadbabaie \emph{et al.} adopt a more conservative
approach, ``which ignores how $\sigma$ depends on the agent
positions in the plane and assumes instead that $\sigma$ might be
any switching signal in some suitably defined set of
interests."\cite{JLM03}

We in this paper follow this basic assumption and formalize the
Vicsek model as follows:

\begin{defi}[Vicsek model]\label{def:naive} {\rm
Given $n$ agents, labeled $1, 2, \cdots, n$, moving in the plane
at discrete time $t\in\nature$, let $\mathcal{P}$ be all simple
undirected graphs over $V=\{1,2,\cdots,n\}$. A Vicsek model is a
pair $(V,\sigma)$, where $\sigma: \nature\ra\mathcal{P}$ is a
switching signal.

For each agent $i$, define $i$'s $\sigma$-neighborhood at time
$t$, written $\mathcal{N}_i(t)$, to be the set of agents that is
connected to $i$ by an edge in graph $\sigma(t)$. That is, agent
$j$ is a neighbor of agent $i$ if and only if $(i,j)$ is an edge
in graph $\sigma(t)$.

Given an initial heading
$\theta(0)=\langle\theta_i(0)\rangle_{i=1}^n$, agent $i$'s heading
$\theta_i(t)$ evolves in discrete time according to
Eq.~\ref{eq:lur}. Namely, agent $i$'s heading at time $t+1$ is the
average of the headings of agent $i$ and its neighbors at time
$t$. }\end{defi}
\begin{remark}\label{rmk:vicsek}
{\rm This definition of a multi-agent coordination model is very
general and more flexible. Several dimensions of
extension/completion could be incorporated in this model: 1) We
can choose either closed/open disk, a triangle-like zone or any
subset of $V$ as the neighborhood; 2) The velocity could also
change in discrete time; 3) We could consider other state
variables of agents besides their headings; and 4) The
neighbor-graph could also be directed. This flexibility would be
helpful in practical applications. }
\end{remark}

The goal of this paper is to show for a large class of Vicsek
models (or switching signals) and for any initial set of agent
headings that the headings of all $n$ agents will converge into
the same heading. Compared with the results obtained in
\cite{JLM03}, ours are more general.

\section{A sufficient condition for multi-agent coordination}
\subsection{Notations and preliminaries}
Suppose $(V,\sigma)$ is a Vicsek model. Following Savkin
\cite{Sav04}, we define a graph $\sigma(\infty)$ over
$V=\{1,2,\cdots,n\}$ as follows: for any two nodes $i,j$, $(i,j)$
is an edge in $\sigma(\infty)$ if and only if for any $K>0$, there
exists some $k\geq K$ such that $(i,j)$ is an edge in graph
$\sigma(k)$.  For convenience, given a collection of graphs
$\{\mathbb{G}_x: x\in X\}$, we write $\biguplus_{x\in X}
\mathbb{G}_x$ for the union of these graphs, \emph{i.e.}, any pair
$(i,j)$ is an edge in $\biguplus_{x\in X} \mathbb{G}_x$ if and
only if it is an edge in some $\mathbb{G}_x$. Then it is easy to
show that there exists some $K>0$ such that
$\sigma(\infty)=\biguplus_{t\geq k}\sigma(t)$ holds for all $k\geq
K$. In this paper we will show that all agents shall eventually
move in the same heading provided that $\sigma(\infty)$ is
connected. This condition is more general than the one given in
\cite{JLM03}, where the authors require that the $\sigma(t)$'s are
periodically jointly connected. In what follows, a switching
signal $\sigma$ is called \emph{finally jointly connected} if
$\sigma(\infty)$ is connected. Clearly this is equivalent to
saying that $\biguplus_{t\geq k}\sigma(t)$ is connected for any
$k\in\nature$.

For a sequence $\{f(k)\}$ and a number $u$ in $\real$, we say $u$
is an \emph{accumulation} point of $\{f(k)\}$ if there is a
subsequence of $\{f(k)\}$ that converges to $u$. We write
$\accu(\{f(k)\})$ for the set of accumulation points of
$\{f(k)\}$.

Given a Vicsek model $(V,\sigma)$ and an initial headings
$\theta(0)=\langle\theta_i(0)\rangle_{i=1}^n$, we now fix some
notations concerning the model.

For $i=1,2,\cdots, n$, define
\begin{eqnarray}
  \Theta_i &=& \{\theta_i(t):t\in\nature\} \\
  m_i &=& \min\accu\Theta_i \\
  M_i &=& \max\accu\Theta_i \\
m &=& \min_{i=1}^n m_i \label{eq:m}\\
M &=& \max_{i=1}^n M_i \label{eq:M}.
\end{eqnarray}
Note that $\Theta_i$  is a bounded set and, therefore, has minimum
and maximum elements.

For any $t\in\nature$, define
\begin{eqnarray}
\underline{\theta}(t)&=&\min_{i=1}^n\theta_i(t)\\
\overline{\theta}(t) &=&\max_{i=1}^n\theta_i(t).
\end{eqnarray}
The following lemma shows $\underline{\theta}(t)\leq m\leq
M\leq\overline{\theta}(t)$.
\begin{lemma}\label{lemma:bound}
For any $t\in\nature$, we have $\underline{\theta}(t)\leq
\underline{\theta}(t+1)\leq m\leq
M\leq\overline{\theta}(t+1)\leq\overline{\theta}(t)$.
Consequently, we have $\lim_{t\ra\infty}\underline{\theta}(t)= m$
and $\lim_{t\ra\infty}\overline{\theta}(t)= M$.
\end{lemma}
\begin{proof}For any non-negative $t$, note that by Vicsek's
nearest-neighbor rule (Eq.~\ref{eq:lur}), we have
$\underline{\theta}(t)\leq\theta_i(t+1)\leq\overline{\theta}(t)$.
In particular, we have
$\underline{\theta}(t)\leq\underline{\theta}(t+1)\leq\overline{\theta}(t+1)\leq
\overline{\theta}(t)$. Now since $\{\underline{\theta}(t)\}$
($\{\overline{\theta}(t)\}$, resp.) is a bounded ascending
(descending, resp.) sequence, it has a limit. We now show its
limit is $m$ ($M$, resp.). Take $\{\underline{\theta}(t)\}$ as an
example. Since it is convergent, any subsequence of
$\{\underline{\theta}(t)\}$ also converges to its limit. Suppose
$\{f(k)\}$ is a sequence such that
$\lim_{k\ra\infty}\theta_i(f(k))=m$ for some agent $i$. Note that
$\underline{\theta}(f(k))\leq\theta_i(f(k))$ for any $k$, we have
$\lim_{t\ra\infty}\underline{\theta}(t)
=\lim_{k\ra\infty}\underline{\theta}(f(k)) \leq
\lim_{k\ra\infty}\theta_i(f(k))=m$. On the other hand, since there
exists some agent $i$ such that
$\{t:\theta_i(t)=\underline{\theta}(t)\}$ is infinite, we have a
sequence $\{g(k)\}$ such that
$\theta_i(g(k))=\underline{\theta}(g(k))$. This shows that
$\lim_{t\ra\infty}\underline{\theta}(t)
=\lim_{k\ra\infty}\underline{\theta}(g(k))
=\lim_{k\ra\infty}\theta_i(g(k))\geq m$ since $m$ is the minimum
accumulation point. As a result, we have
$\lim_{t\ra\infty}\underline{\theta}(t)=m$. Similarly, we can show
$\lim_{t\ra\infty}\overline{\theta}(t)=M$. So we have
$\underline{\theta}(t)\leq m\leq M\leq\overline{\theta}(t)$ for
any $t$.
\end{proof}
Note that as shown in the proof of the above lemma, we have a
sequence, say $\{f(k)\}$, such that
$\theta_i(f(k))=\underline{\theta}(f(k))$ and
$\lim_{k\ra\infty}\theta_i(f(k))= m$ for some agent $i$.
Similarly, we have a sequence, say $g(k)$, such that
$\lim_{k\ra\infty}\theta_j(g(k))= M$ for some $j$.
\subsection{Leaderless coordination}
\begin{thm}\label{thm:leaderless}
Given a Vicsek model $(V,\sigma)$, suppose $\sigma(\infty)$ is
connected. Then for any
$\theta(0)=\langle\theta_i(0)\rangle_{i=1}^n$, we have
\begin{equation}\label{eq:limit}
\lim_{t\ra\infty}\theta_i(t)=\theta_{ss}\ \ (i=1,2,\cdots,n)
\end{equation}
where $\theta_{ss}$ is a number depending only on $\theta(0)$ and
$\sigma$.
\end{thm}

To prove this theorem, we need several lemmas.

Recall $V=\{1,2,\cdots,n\}$. For a graph $\mathbb{G}$ over $V$ and
any two disjoint subsets $A$, $B$ of $V$, we say $A$ and $B$ are
\emph{connected} if there exist $a\in A, b\in B$ such that $(a,b)$
is an edge in $\mathbb{G}$. If $A$ happens to be a singleton
$\{a\}$, we also say node $a$ is connected to $B$. In this case,
we say alternatively $a$ has a neighbor in $B$.

The following lemma suggests that, if the agents are divided into
two parts such that the maximum heading of the first part is
sufficiently smaller than the minimum of the second part, then,
after updating the headings using Eq.~\ref{eq:lur}, the agents
will also form two parts such that one part is still sufficiently
smaller than the rest.

For $a<b$ in $\real$ and any natural number $t$, we write
$V_t(a,b)=\{i\in V: a<\theta_i(t)<b\}$.

\begin{lemma}\label{lemma:key}
Given $\alpha<\beta<\gamma$ and set $\delta=\beta-\alpha$,
$\e=\delta/n^n$, suppose $V_t(\alpha-\e,\alpha+\e)$ and
$V_t(\beta-\e,\gamma+\e)$ are two nonempty disjoint
subsets of $V$ such that their union is $V$. We have\\
{\rm(1)} If $V_t(\alpha-\e,\alpha+\e)$ and
$V_t(\beta-\e,\gamma+\e)$ are disconnected at time $t$, then
$V_{t+1}(\alpha-\e,\alpha+\e)=V_t(\alpha-\e,\alpha+\e)$ and
$V_{t+1}(\beta-\e,\gamma+\e)=V_t(\beta-\e,\gamma+\e)$.\\
{\rm(2)} If $V_t(\alpha-\e,\alpha+\e)$ and
$V_t(\beta-\e,\gamma+\e)$ are connected at time $t$, then
$V_{t+1}(\alpha-\e,\alpha+\e)=V_t(\alpha-\e,\alpha+\e)-\{i\in V: i
\mbox{\rm\ has\ a\ neighbor\ in}\ V_t(\beta-\e,\gamma+\e)
\mbox{\rm\ at\ time}\ t\}$ and
$V-V_{t+1}(\alpha-\e,\alpha+\e)=V_{t+1}(\alpha+\delta/n-\e,\gamma+\e)$.
\end{lemma}
\begin{proof} If $V_t(\alpha-\e,\alpha+\e)$
and $V_t(\beta-\e,\gamma+\e)$ are disconnected at time $t$, then
for any $i\in V_t(\alpha-\e,\alpha+\e)$, its neighbors are all in
$V_t(\alpha-\e,\alpha+\e)$. By Eq.~\ref{eq:lur}, we have
$\alpha-\e<\theta_i(t+1)< \alpha+\e$. Similarly, for any $j\in
V_t(\beta-\e,\gamma+\e)$, we have $\beta-\e<\theta_j(t+1)<
\gamma+\e$.

On the other hand, suppose $V_t(\alpha-\e,\alpha+\e)$ and
$V_t(\beta-\e,\gamma+\e)$ are connected at time $t$. For $i\in
V_t(\alpha-\e,\alpha+\e)$, if all its neighbors are in
$V_t(\alpha-\e,\alpha+\e)$, then $i\in
V_{t+1}(\alpha-\e,\alpha+\e)$; if $i$ has a neighbor, say $j_0$,
in $V_t(\beta-\e,\gamma+\e)$, then we have
\begin{eqnarray*}
  \theta_i(t+1) &=&
    \frac{\theta_i(t)+\Sigma_{j\in\mathcal{N}_i(t)}\theta_j(t)}
    {1+n_i(t)}\\
   &=&\frac{\theta_i(t)+\theta_{j_0}(t)+
   \Sigma_{j\in\mathcal{N}_i(t),j\not=j_0}\theta_j(t)}
    {1+n_i(t)}\\
   &>&\frac{(\alpha-\e)+(\beta-\e)+ (n_i(t)-1)\times (\alpha-\e)}{1+n_i(t)}  \\
   &=&\frac{\beta-\alpha  + (1+n_i(t))\times (\alpha-\e)}{1+n_i(t)} \\
   &=&(\alpha-\e)+\frac{\beta-\alpha}{1+n_i(t)}\\
   &\geq&\alpha-\e+\delta/n.
\end{eqnarray*}
Note that $\theta_i(t+1)<\gamma+\e$ holds for any $i\in V$. This
shows that, if $i\in V_t(\alpha-\e,\alpha+\e)$ has a neighbor in
$V_t(\beta-\e,\gamma+\e)$, then $i\in
V_{t+1}(\alpha+\delta/n-\e,\gamma+\e)$. Similarly, for any $j\in
V_t(\beta-\e,\gamma+\e)$, we can  show
$\theta_j(t+1)>\alpha+\delta/n-\e$. In summary, we have $i\in
V_{t+1}(\alpha-\e,\alpha+\e)$ if and only if $i\in
V_t(\alpha-\e,\alpha+\e)$ and it has a neighbor in
$V_t(\beta-\e,\gamma+\e)$ at time $t$. As for any other agent $j$,
we have $j\in V_{t+1}(\alpha+\delta/n-\e,\gamma+\e)$.
\end{proof}

\begin{lemma}\label{lemma:converge}
Suppose $\sigma(\infty)$ is connected and $\{f(k)\}$ is a
sequence. Then we have a subsequence $\{g(k)\}$ of $\{f(k)\}$ such
that all $\{\theta_i(g(k))\}$ are convergent for $i\in V$.
\end{lemma}
\begin{proof}This follows from the compactness of $[0,2\pi]$ and
that $\theta_i(t)\in [0,2\pi)$ for any $i$, $t$.
\end{proof}

\begin{lemma}\label{lemma:mM}
Suppose $\{g(k)\}$ is a sequence such that $\{\theta_i(g(k))\}$
converges to $l_i$ for $i=1,2,\cdots,n$. Then $m=\min_{i=1}^n l_i$
and $M=\max_{i=1}^n l_i$, where
$m=\lim_{t\ra\infty}\underline\theta(t)$,
$M=\lim_{t\ra\infty}\overline\theta(t)$ and
$\underline\theta(t)=\min_{i=1}^n\theta_i(t)$,
$\overline\theta(t)=\max_{i=1}^n\theta_i(t)$.
\end{lemma}
\begin{proof}
 Take $m=\min_{i=1}^n l_i$ as an
example. Note that there exists some $i$ such that $\{k:
\theta_i(g(k))=\underline{\theta}(g(k))\}$ is infinite. We have a
subsequence $\{h(k)\}$ of $\{g(k)\}$ such that
$\theta_i(h(k))=\underline{\theta}(h(k))$ and
$l_i=\lim_{k\ra\infty}\theta_i(h(k))=\lim_{k\ra\infty}\underline{\theta}(h(k))=m$.
That $M=\max_{i=1}^n l_i$ is similar.
\end{proof}

\begin{proof}[Proof of Theorem~\ref{thm:leaderless}]
To show that these autonomous agents shall eventually move into
the same heading, we need only to show $m=M$.  We prove this by
reduction to absurdity.

Suppose $m<M$ and $\{g(k)\}$ is a sequence such that
$\{\theta_i(g(k))\}$ converges to $l_i$ for $i=1,2,\cdots,n$.
Recall by Lemma~\ref{lemma:mM} that $m=\min_{i=1}^n l_i$ and
$M=\max_{i=1}^n l_i$. Set $l=\min\{l_i:l_i>m\}$ and take
$\delta=l-m$, $\e=\delta/n^n$. Then there exists $K>0$ such that
$\theta_i(g(k))\in(l_i-\e,l_i+\e)$ for $k\geq K$ and
$i=1,2,\cdots,n$. Moreover, we have $V_{g(k)}(m-\e,m+\e)=\{i\in V:
l_i=m\}$ and $V_{g(k)}(l-\e,M+\e)=\{i\in V: l_i\geq l\}$. Clearly
$V_{g(k)}(m-\e,m+\e)$ and $V_{g(k)}(l-\e,M+\e)$ satisfy the
condition of Lemma~\ref{lemma:key}.

Now since $\biguplus_{t\geq p}\sigma(t)$ is connected for any
$p\in\nature$, we have some $k\geq K$ such that
$V_{g(k)}(m-\e,m+\e)$ and $V_{g(k)}(l-\e,M+\e)$ are connected at
time $g(k)+w$ for some $0\leq w< g(k+1)-g(k)$. Fix one such $k$
and suppose $g(k+1)-g(k)=W$. For each $w=0,1,\cdots, W$, we define
$A_w=V_{g(k)+w}(m-\e,m+\e)$ and $B_w=V-A_w$.

Note that because
$A_W=V_{g(k+1)}(m-\e,m+\e)=V_{g(k)}(m-\e,m+\e)=A_0$, it is also
true that
$m-\e<\underline{\theta}(g(k))\leq\underline{\theta}(g(k+1))<m+\e$.
Recall that because $\underline{\theta}(t)$ is an ascending chain
(see Lemma~\ref{lemma:bound}), we also have
$\underline{\theta}(g(k)+w)\in(m-\e,m+\e)$ for any $w=1,2,\cdots,
W-1$.

Set $C=\{w\in[0,W): A_w \mbox{\rm\ and}\ B_w \mbox{\rm\ are\
connected\ at\ time}\ g(k)+w\}$. Clearly $C$ is not empty since
there exists some $w$ such that $A_0$ is connected to $B_0$ at
time $g(k)+w$. Suppose $C=\{w_1,w_2,\cdots,w_q\}$ and $0\leq w_1<
w_2 <\cdots < w_q < W$. We claim
\begin{eqnarray}\label{eq: chain1}
A_0\supsetneq A_{w_1+1}\supsetneq A_{w_2+1}\supsetneq \cdots
\supsetneq A_{w_{q-1}+1}\supsetneq A_{w_q+1} \\
\label{eq: chain2}
B_{w_s+1}=V_{g(k)+w_s+1}(m+\frac{\delta}{n^s}-\e, M+\e)\
(s=1,2,\cdots,q)
\end{eqnarray}

As the induction basis, note that $A_0=V_{g(k)}(m-\e,m+\e)=\{i:
l_i=m\}$ and $B_0=\{i:l_i\geq
l\}=V_{g(k)}(l-\e,M+\e)=V_{g(k)}(m+\delta-\e,M+\e)$.

Note that $w_1$ is the first index $w$ such that $A_w$ is
connected to $B_w$; by Lemma~\ref{lemma:key}, we have $A_0=A_w$
and $B_w=V_{g(k)+w}(m+\delta-\e,M+\e)$ for any $w\leq w_1$.
Moreover, since $A_{w_1}=A_0$ is connected to $B_{w_1}=B_0$ at
time $g(k)+w_1$, by Lemma~\ref{lemma:key}, we have
\begin{eqnarray}\label{eq:chain10}
A_{w_1+1}=A_{w_1}-\{i\in A_{w_1}: i \mbox{\rm\ has\ a\ neighbor\
in}\ B_{w_1} \mbox{\rm\ at\ time}\ g(k)+w_1\}\subsetneq A_0\\
\label{eq:chain20}
B_{w_1+1}=V-A_{w_1+1}=V_{g(k)+w_1+1}(m+\delta/n-\e,M+\e)
\end{eqnarray}

Recall that $A_{w_1+1}\not=\emptyset$ since
$\underline\theta(g(k)+w_1+1)\in (m-\e,m+\e)$.

Suppose for $s<q$ we have
\begin{eqnarray}\label{eq:chain1s}
A_0\supsetneq A_{w_1+1}\supsetneq A_{w_2+1}\supsetneq \cdots
\supsetneq A_{w_{s-1}+1}\supsetneq A_{w_s+1}\\
\label{eq:chain2s}
B_{w_j+1}=V_{g(k)+w_j+1}(m+\frac{\delta}{n^j}-\e, M+\e)\
(j=1,2,\cdots,s)
\end{eqnarray}
Note that $s<n-1$ must hold since $A_0$ contains at most $n-1$
agents and $A_0\supsetneq A_{w_1+1}\supsetneq\cdots\supsetneq
A_{w_{s}+1}\not=\emptyset$.

We now show $A_{w_{s}+1}\supsetneq A_{w_{s+1}+1}$ and
$B_{w_{s+1}+1}=V_{g(k)+w_{s+1}+1}(m+\frac{\delta}{n^{s+1}}-\e,
M+\e)$.

Note that $w_{s+1}$ is the first index $w>w_s$ such that $A_w$ is
connected to $B_w$. By Lemma~\ref{lemma:key}, we have
$A_w=A_{w_s+1}$ and
$B_w=V_{g(k)+w}(m+\frac{\delta}{n^{s}}-\e,M+\e)=B_{w_s+1}$ for any
$w\in (w_s,w_{s+1}]$. Moreover, since $A_{w_{s+1}}=A_{w_s+1}$ is
connected to $B_{w_{s+1}}=B_{w_s+1}$ at time $g(k)+w_{s+1}$, by
Lemma~\ref{lemma:key}, we have
\begin{eqnarray}\label{eq:chain1s1}
A_{w_{s+1}+1}=A_{w_{s+1}}- {\hskip 100mm\nonumber}\\
\{i\in A_{w_{s+1}}: i \mbox{\rm\ has\
a\
neighbor\ in}\ B_{w_{s+1}} \mbox{\rm\ at\ time}\ g(k)+w_{s+1}\}\subsetneq A_{w_s+1}\\
\label{eq:chain2s1}
B_{w_{s+1}+1}=V-A_{w_{s+1}+1}=V_{g(k)+w_{s+1}+1}(m+\frac{\delta}{n^{s+1}}-\e,M+\e)
\end{eqnarray}
In summary, we have obtained that $A_0=A_{w_1}\supsetneq
A_{w_q+1}$.

Note that if $w_q<W-1$, then $A_w$ and $B_w$ are disconnected for
any $w\in(w_q, W)$. By Lemma~\ref{lemma:key} again, we know
$A_w=A_W$ for $w\in(w_q,W]$. In particular, we have
$A_{w_q+1}=A_W$. On the other hand, if $w_q=W-1$, we also have
$A_{w_q+1}=A_W$.

This suggests that if $m<M$, then $A_0\not=A_W$. This is a
contradiction. So our assumption that $m<M$ cannot hold. This ends
the proof of this theorem.
\end{proof}

\begin{remark}{\rm
Note that if $\sigma:\nature\ra\mathcal{P}$ is a switching signal
for which there exists an infinite sequence of bounded,
non-overlapping (but not necessarily contiguous) intervals across
which the $n$ agents are linked together, then $\sigma(\infty)$ is
connected. By the above theorem, we know all agents would
eventually move in the same heading for this $\sigma$.
Consequently, this theorem shows the desired condition given in
\cite[p990, below Theorem 2]{JLM03} is a sufficient condition for
asymptotic convergence.

The hypothesis of Theorem~\ref{thm:leaderless}, however, is still
not necessary. For example, if some $\sigma(t)$ is the complete
graph over $V$, then a coordination could be achieved at time
$t+1$. But if it is not connected, $\sigma(\infty)$ will have
$1<p\leq n$ connected components, say
$\mathbb{G}_1,\mathbb{G}_2,\cdots,\mathbb{G}_p$. Similar to the
argument given above for Theorem~\ref{thm:leaderless}, we can show
for any $h=1,2,\cdots,p$, there exists a heading $\hat{\theta}_h$
such that $\lim_{t\ra\infty}\theta_i(t)=\hat{\theta}_h$ for any
$i\in\mathbb{G}_h$.}
\end{remark}

\subsection{Leader-following coordination}
In \cite{JLM03}, Jadbabaie \emph{et al.} also consider a modified
version of Vicsek's discrete-time system, which consists of the
same group of $n$ agents as before except that one leader agent,
labeled 0, is added. Agent 0 moves at the same constant speed as
its $n$ followers but with a fixed heading $\theta_0$. Agent $i$
then updates its heading using the average of its own heading plus
the headings of its neighbors. Note that this time the leader may
be in its neighborhood.

Our abstract Vicsek  model with a leader now can be formulated as
follows:
\begin{defi}\label{def:leader}{\rm
Suppose $V^+=\{0,1,\cdots,n\}$ and $\mathcal{P}^+$ is the
collection of simple undirected graphs over $V^+$. A
leader-following Vicsek model is just a pair $(V^+,\sigma)$, where
$\sigma:\nature\ra\mathcal{P}^+$ is a switching signal.

For each agent $i>0$, define $i$'s $\sigma$-neighborhood at time
$t$, written $\mathcal{N}_i(t)$, to be the set of agents that are
connected to $i$ by an edge in the graph $\sigma(t)$. That is,
agent $j$ is a neighbor of agent $i$ if and only if $(i,j)$ is an
edge in the graph $\sigma(t)$.

Given an initial heading
$\theta(0)=\langle\theta_i(0)\rangle_{i=1}^n$ and a fixed heading
$\theta_0$ in which agent 0 moves at all times, for $i>0$, agent
$i$'s heading evolves in discrete time according to the following
equation:
\begin{equation}\label{eq:leader}
\theta_i(t+1)=
\frac{\theta_i(t)+\Sigma_{j\in\mathcal{N}_i(t)}\theta_j(t)}
{1+n_i(t)}
\end{equation}
where $n_i(t)$ is the number of agents in $\mathcal{N}_i(t)$.}
\end{defi}

For a leader-following Vicsek model, we have the following
correspondence of Theorem~\ref{thm:leaderless}.
\begin{thm}\label{thm:leader}
Given a leader-following Vicsek model $(V^+,\sigma)$, suppose
$\sigma(\infty)$ is connected. Then for any
$\theta(0)=\langle\theta_i(0)\rangle_{i=1}^n$ and $\theta_0$, we
have $\lim_{t\ra\infty}\theta_i(t)=\theta_0$ for all
$i=1,2,\cdots,n$.
\end{thm}
\begin{proof}Note that Lemma~\ref{lemma:key} and Lemma~\ref{lemma:converge}
still hold for the leader-following case; this theorem follows
from a similar argument as given for Theorem~\ref{thm:leaderless}.
\end{proof}

\section{Conclusions and further work}
In \cite{JLM03}, Jadbabaie \emph{et al.} show that all agents
shall move in the same heading if the neighbor-graphs are
\emph{periodically jointly connected}, \emph{i.e.}, the union of
any $T$ contiguous neighbor-graphs is connected for some fixed
$T$. In the same paper, they also ask whether this still holds
when there exists a sequence of uniformly bounded intervals over
the discrete time such that the union of neighbor-graphs across
each interval is connected?

This paper has shown that all agents shall move in the same
heading under a very weak condition that requires the
neighbor-graphs to be \emph{finally jointly connected},
\emph{i.e.}, the union of all graphs started from any time is
connected. This result gives an affirmative answer to the question
raised in \cite{JLM03}.

It should be emphasized that results obtained in this paper are
valid for many versions of the Vicsek model (or coordination
multi-agent models that use nearest-neighbor rule to update their
state) (see Remark~\ref{rmk:vicsek}). As for some specific
versions of the Vicsek model, there have been some impressive
results. Recently, Jadbabaie \cite{Jad03} has shown that, if we
choose the neighborhood region to be open, then a necessary and
sufficient condition for all headings to converge to the same
heading is that the neighbor-graph does not change after finite
steps and is connected.\footnote{A similar result was obtained by
Savkin \cite{Sav04} for a simple version of Vicsek model where the
headings $\theta_i(t)$'s take their values from a certain finite
set. } Jadbabaie also notes that \cite[p.8, last paragraph]{Jad03}
the problem would be more complicated if a closed neighborhood
region were chosen. It seems this method cannot be directly
applied to other kinds of neighborhoods. As a matter of fact,
there are often situations when agents do not have disk-like
visibility but rather might have a cone-like field of
view.\footnote{Such a situation has been investigated (though for
continuous time) by Lin \emph{et al.} \cite{LBF04}.}

Note that our results are based on the assumption that the
switching signal is pre-specified. In our future work, we shall
plan to develop a model that can explain how the neighbor-graphs
evolve over discrete time, and determine sufficient conditions for
coordination of multi-agents in terms of these agents' initial
states. Another thing that should be stressed is that coordination
results obtained in ours and in \cite{JLM03} are all asymptotic.
It will be interesting to devise other local updating rules such
that, using these rules, coordination will be reached quickly and
without centralized control. This work is currently being
undertaken.


\bibliography{vicsek}
\end{document}